\documentclass[draft]{agujournal2019}
\usepackage{color}
\usepackage{url} 
\usepackage{lineno}

\usepackage{amsmath}
\usepackage{amsfonts}
\usepackage{amssymb}

\draftfalse

\newcommand{\ssr}{    {Space Sci. Rev.}}
\newcommand{\grl}{    {Geophys. Res. Lett.}}
\newcommand{\jgr}{    {J. Geophys. Res.}}

\newcommand{\apj}{    {The Astrophysical Journal}}
\newcommand{\blue}{\textcolor{black}}

\journalname{Geophysical Research Letters}

\begin{document}

%
%





\title{Configuration of magnetotail current sheet prior to magnetic reconnection onset}

\authors{Xin An\affil{1}, Anton Artemyev\affil{1,2}, Vassilis Angelopoulos\affil{1}, Andrei Runov\affil{1}, San Lu\affil{3}, Philip Pritchett\affil{4}}


\affiliation{1}{Department of Earth, Space and Planetary Sciences, University of California, Los Angeles, CA, 90095, USA}
\affiliation{2}{Space Research Institute of the Russian Academy of Sciences, Moscow, 117997, Russia}
\affiliation{3}{School of Earth and Space Sciences, University of Science and Technology of China, Hefei, 230026, China.}
\affiliation{4}{Department of Physics and Astronomy, University of California, Los Angeles, CA, 90095, USA}

\correspondingauthor{Xin An}{xinan@epss.ucla.edu}

\begin{keypoints}
\item Nongyrotropic pressure gradients alter significantly the configuration of thin magnetotail current sheets before reconnection onset.
\item Ion pressure nongyrotropy is caused by velocity shear of off-equatorial ion outflow during current sheet thinning.
\item Nongyrotropic pressure may be critical to the formation and stability of thin current sheets in the magnetotail.
\end{keypoints}

\begin{abstract}
The magnetotail current sheet configuration determines magnetic reconnection properties that control the substorm onset, one of the most energetic phenomena in the Earth's magnetosphere. The quiet-time current sheet is often approximated as a two-dimensional (2D) magnetic field configuration balanced by isotropic plasma pressure gradients. However, reconnection onset is preceded by the current sheet thinning and the formation of a nearly one-dimensional (1D) magnetic field configuration. In this study, using particle-in-cell simulations, we investigate the force balance of such thin current sheets when they are driven by plasma inflow. We demonstrate that the magnetic field configuration transitions from 2D to 1D thanks to the formation of plasma pressure nongyrotropy and reveal its origin in the nongyrotropic terms of the ion distributions. We show that substorm onset may be controlled by the instability and dynamics of such nongyrotropic current sheets, having properties much different from the most commonly investigated 2D isotropic configuration.
\end{abstract}

\section*{Plain Language Summary}
The geomagnetic field lines are stretched by the solar wind into an elongated magnetotail behind the Earth, a region supported by strong equatorial sheet-like plasma currents. The stability of the magnetotail current sheet configuration controls the substorm onset, one of the most powerful phenomena in the Earth's magnetosphere. The most commonly investigated current sheet models based on the isotropic stress balance have difficulties in describing the most intense thin current sheets. Using computer simulations, we find that thinning of the magnetotail current sheet creates a quasi-1D equilibrium accompanied by ion pressure nongyrotropy. Therefore, substorm onset may be controlled by the instability and dynamics of such nongyrotropic current sheets, having properties much different from the most commonly investigated 2D isotropic configuration.

\section{Introduction}
The magnetospheric substorm is a very energetic phenomenon associated with magnetic energy release, particle acceleration and injection into the inner magnetosphere, and energy dissipation via magnetosphere-ionosphere coupling \cite{Baker96, Angelopoulos08, Sitnov19}. \blue{According to the dominant concept, substorm} onset is determined by magnetic reconnection in the middle magnetotail current sheet \cite{Nagai&Machida98,Petrukovich98,Angelopoulos13,Torbert18,Lu20:natcom}. The reconnection can result from either driven \cite{Pritchett91,Pritchett05:driven,Sitnov09,Liu14:CS,Pritchett&Lu18} or spontaneous \cite{Sitnov13,Sitnov17} current sheet instabilities, which have been the focus of theoretical investigations starting from first publications on the tearing mode \cite{Coppi66,Schindler74,Galeev76} and up to date \cite<see reviews by>[and references therein]{Lui04, Sitnov19}. The dominant concept supported by spacecraft observations \cite<e.g.,>[]{Petrukovich07,Sergeev11,Artemyev16:jgr:thinning} and numerical simulations \cite<e.g.,>[]{Lu18:3dthinning} is that external driver thins the magnetotail current sheet and such thin current sheet becomes unstable to magnetic reconnection. Although questions about the threshold thickness and the reconnection onset have not yet been resolved \cite<see discussion in, e.g.,>[]{Sitnov21}, the current sheet thinning is well documented in spacecraft data \cite<see reviews by>[and references therein]{Petrukovich15:ssr,Runov21:jastp}, numerical simulations \blue{\cite{Pritchett&Coroniti94,Pritchett&Coroniti95,Hesse01,Liu14:CS,Hsieh&Otto15}} and empirical models \cite{Stephens19,Tsyganenko21}. The investigation of the reconnection onset requires detailed information on the pre-onset current sheet configuration, which is the focus of our study.

The most investigated class of thin current sheet models consists of various 2D plasma equilibria with the equatorial current density $J_y$ (the Geocentric Solar Magnetospheric coordinate is used throughout the paper) constrained by the plasma pressure gradient $\partial P/\partial x $ and the equatorial magnetic field $B_z$: $J_y=c(\partial P/\partial x)/B_z$ \cite{bookSchindler06}. Two representative models from this class include one with a monotonic profile of $B_z(x)$ \cite{Schindler72,LP82} and another one with a local maximum of $B_z(x)$ \cite{Sitnov10}. The first model has been proven to be quite stable to magnetic reconnection \cite<see>[]{Pellat91,Quest96}, and thus an external driver should be sufficiently strong to thin the current sheet almost to electron scales before reconnection can occur \cite{Pritchett05:driven,Hesse01,Liu14:CS,Lu20:natcom}. The second model may be unstable to magnetic reconnection even for an ion-scale current sheet \cite{Sitnov13,Pritchett15:pop,Bessho&Bhattacharjee14}. However, both models have difficulties in describing the most intense thin current sheets, those with $J_y>c(\partial P/\partial x)/B_z$ often observed in the magnetotail \cite{Artemyev16:jgr:pressure,Artemyev21:grl} and even in the reconstructed magnetotail empirical models \cite<see discussion in>[]{Sitnov21}. If the current density $J_y$ exceeds $c(\partial P/\partial x)/B_z$, the magnetic and thermal stress balance in the magnetotail current sheet requires a contribution from pressure anisotropy \cite<e.g., electron anisotropy, see>[]{Artemyev16:pop:cs} or nongyrotropy \cite<e.g., $P_{xz} \neq 0$ due to ion demagnetized motion around the equatorial plane; see>[]{Burkhart92TCS,Maha94,Mingalev07}. Electron anisotropy can be quite strong for some of the observed current sheets \cite<see examples in>[]{Lu19:jgr:cs,Kamaletdinov20}, but cannot balance more than $30$\% of $J_y$ for the majority of thin current sheets \cite{Artemyev20:jgr:electrons}. The ion nongyrotropy has not been well measured \cite<see discussion in>[]{Aunai11:angeo,Artemyev19:jgr:ions}, but theoretically it could potentially be sufficiently strong to balance the observed current density magnitudes as $J_y\sim c(\partial P_{xz}/\partial z)/B_z$ \cite<see discussion in>[]{Zhou09,Artemyev&Zelenyi13,Sitnov&Merkin16}.

In this Letter we investigate a fine structure of the thin current sheet that forms right before the externally driven reconnection of a 2D equilibrium which initially obeys $J_y=c(\partial P/\partial x)/B_z$ \cite{LP82}. Thinning of such a 2D current sheet may change the magnetic field configuration and create a new quasi-1D equilibrium accompanied by ion pressure nongyrotropy. We start with a polarized current sheet where electron currents exceed ion currents \cite{Lu19:apj,Sitnov21:grl}. Such polarization of thin current sheets has been demonstrated theoretically \cite{Hesse98, SB02,Zelenyi10GRL,Lu16:cs} and has been suggested in order to explain the dominance of electron currents in spacecraft observations of current sheets \cite<see, e.g.,>[]{Runov06,Artemyev09:angeo}. Tracing the evolution of stress balance terms ($J_y B_z/c$, $\partial P/\partial x$, and $\partial P_{xz}/\partial z$ for ions and electrons), we demonstrate that before the reconnection onset the current sheet configuration becomes almost 1D with $J_y B_z/c \approx \partial P_{xz}/\partial z\gg \partial P/\partial x$, where $\partial P_{xz}/\partial z$ is mostly due to ions and $J_y$ is mostly carried by electrons due to strong ${\bf E}\times{\bf B}$ drift. We also investigate the formation of this finite $P_{xz}$, and show the main origin of the pressure nongyrotropy.

\section{Computational setup}
\subsection{Initial equilibrium}
The initial equilibrium is identical to the one used in our previous studies \cite{an2022suppression,an2022fast}. Here we briefly recap how such equilibrium was obtained. We generalize the Lembege-Pellat current sheet \cite{LP82} to include the background plasma and the polarization electric field commonly observed in magnetotail current sheet \cite{Vasko14:angeo,Lu19:jgr:cs}. This polarization electric field is caused by the decoupling between ion and electron orbits near the equator $z=0$, leading to stronger electron currents than ion currents \cite{Runov06,Artemyev09:angeo}. The magnetic field lines in the $x$-$z$ plane are described by the vector potential $A_y(\varepsilon x, z) \hat{\mathbf{e}}_y$, and the polarization electric field is derived from the electrostatic potential $\varphi(\varepsilon x, z)$, where $\vert \varepsilon \vert \ll 1$ indicates a weakly nonuniformity in the $x$ direction. The vector and electrostatic potentials are self-consistently determined by Ampere's Law
\begin{linenomath*}
\begin{equation}
    \frac{\partial^2}{\partial z^2} A_y = - 4 \pi \sum_{\alpha} q_\alpha n_0 \frac{v_{D \alpha}}{c} \exp\left( -\frac{q_\alpha \varphi}{T_{\alpha 0}} + \frac{v_{D\alpha} q_\alpha A_y}{c T_{\alpha 0}} \right) , \label{eq:Ay-poisson-boltzmann}
\end{equation}
\end{linenomath*}
and the quasi-neutrality condition
\begin{linenomath*}
\begin{equation}
    \sum_{\alpha} q_\alpha n_0 \exp\left( -\frac{q_\alpha \varphi}{T_{\alpha 0}} + \frac{v_{D \alpha} q_\alpha A_y}{c T_{\alpha 0}} \right) + q_\alpha n_b \exp\left( -\frac{q_\alpha \varphi}{T_{\alpha b}} \right) = 0 , \label{eq:quasi-neutrality}
\end{equation}
\end{linenomath*}
where two populations, including the current sheet population (i.e., the current-carrying one) and the background population (i.e., the non-current-carrying one), are present. Here $c$ is the speed of light, $q_\alpha$ is the charge, $n_0$ is the current sheet density, $n_b$ is the background density, and $v_{D \alpha}$ is the drift velocity, $T_{\alpha 0}$ is the current sheet temperature, and $T_{\alpha b}$ is the background temperature, where the subscript $\alpha = e, i$ stands for electrons and ions, respectively. These parameters are chosen as $n_b / n_0 = 0.2$, $v_{D i} = v_\mathrm{A} / 3$, $v_{D e} = -5 v_\mathrm{A} / 3$, $T_{i b} = T_{i 0} = 5 m_i v_\mathrm{A}^2 / 12$ and $T_{e b} = T_{e 0} = m_i v_\mathrm{A}^2 / 12$, where $m_i$ is the ion mass, $v_\mathrm{A} = B_0 / \sqrt{4 \pi n_0 m_i}$ is the Alfv\'en velocity, and $B_0$ is the reference magnetic field. \blue{The respective current densities and number densities in Equations \eqref{eq:Ay-poisson-boltzmann} and \eqref{eq:quasi-neutrality} are obtained by integrating the drifting Maxwellian \cite{LP82}
\begin{linenomath*}
\begin{equation}
    f_\alpha (x, z, \mathbf{v}) \propto \exp \left( { - \frac{{q_\alpha  }}{{T_{\alpha 0} }}\left( {\varphi  - \frac{{v_{D\alpha } }}{c}A_y } \right) - \frac{{m_\alpha  }}{{2 T_{\alpha 0} }}\left( {{\bf v} - v_{D\alpha } {\bf e}_y } \right)^2 } \right)
\end{equation}
\end{linenomath*}
in velocity space.} We solve Equations \eqref{eq:Ay-poisson-boltzmann} and \eqref{eq:quasi-neutrality} in the rectangular domain $[-L_z / 2 \leqslant z \leqslant L_z / 2] \times [-L_x \leqslant x \leqslant 0]$ with the boundary condition 
\begin{linenomath*}
\begin{equation}
    A_y \big\vert_{z = 0} = \varepsilon B_0 x,\,\, \partial A_y / \partial z \big\vert_{z = 0} = 0 ,
\end{equation}
\end{linenomath*}
where $\varepsilon B_0$ gives the $z$ component of the magnetic field at $z = 0$. Once $A_y$ and $\varphi$ are obtained, we have the electromagnetic fields and density profiles of the equilibrium to initialize the particle-in-cell (PIC) simulation.

\subsection{Kinetic implementation}
We simulate driven magnetic reconnection starting from the above equilibrium using a PIC code developed by \citeA{Pritchett01,Pritchett05:driven}. Our simulation has two dimensions ($\partial / \partial y = 0$) in configuration space and three in velocity space. Normalizations are based on ions: time is normalized to the inverse of the ion cyclotron frequency $\omega_{ci} = e B_0 / (m_i c)$, lengths to the ion inertial length $d_i = c \sqrt{m_i / (4 \pi n_0 e^2)}$, velocities to  $v_\mathrm{A}$, and energies to $m_i v_\mathrm{A}^2$. \blue{The domain size is $L_z \times L_x = 512 \times 2048\, \delta^2$ with a grid size $\delta = d_i / 32$, which gives the range of domain $[-8 \leqslant z / d_i \leqslant 8] \times [-64 \leqslant x / d_i \leqslant 0]$.} The Debye length is $\lambda_{De} = 0.46\, \delta$. The ion-to-electron mass ratio is $m_i / m_e = 100$. The normalized speed of light is $c / v_\mathrm{A} = 20$. The time step is $\Delta t = 0.001\, \omega_{ci}^{-1}$. The reference density $n_0$ is represented by 425 particles per cell, which yields a total of $2 \times 10^8$ particles initially, and approximately this many throughout the run time.

The Lembege-Pellat current sheet with $B_z \neq 0$ is stable to spontaneous reconnection \cite{Pellat91}. Thus we drive a plasma inflow by applying an external electric field at the boundaries $z = \pm L_z / 2$, given by
\begin{linenomath*}
\begin{equation}
    E_{y, \mathrm{drive}} (x, t) = E_{y0} \cdot f(t) \cdot \sin^2\left(\frac{\pi x}{L_x}\right),
\end{equation}
\end{linenomath*}
where the magnitude is controlled by $cE_{y0} / (v_\mathrm{A} B_0) = 0.4$ and $f(t) = \tanh(\omega t) / \cosh^2(\omega t)$ describes the turn-on and turn-off of the applied electric pulse over the characteristic time scale $\omega^{-1} = 20\, \omega_{ci}^{-1}$. New particles are injected into the system at $z = \pm L_z / 2$ at a flux level determined by the background density and the $\mathbf{E} \times \mathbf{B}$ drift velocity. Particles crossing the $z$ boundaries are reflected back into the system.

At $x = -L_x$ and $x = 0$, particles are removed from the system at one cell inside the boundaries. New particles are continuously injected back into the system at those locations with a density profile identical to that of the initial current sheet. The velocities of these newly injected particles are distributed according to a one-sided Maxwellian \cite{aldrich1985particle}. The guard values of $B_z$ at the $x$ boundaries that are used to advance $E_y$ in the Ampere's Law are determined using the two-level time advancement scheme \cite{Pritchett05:driven} in which it is assumed that the $B_z$ field propagates at the speed of light. This scheme ensures that magnetic fluxes exit the $x$ boundaries properly.

\section{Results}
\subsection{Simulation overview}
\begin{figure*}[tphb]
    \centering
    \includegraphics[width=0.6\textwidth]{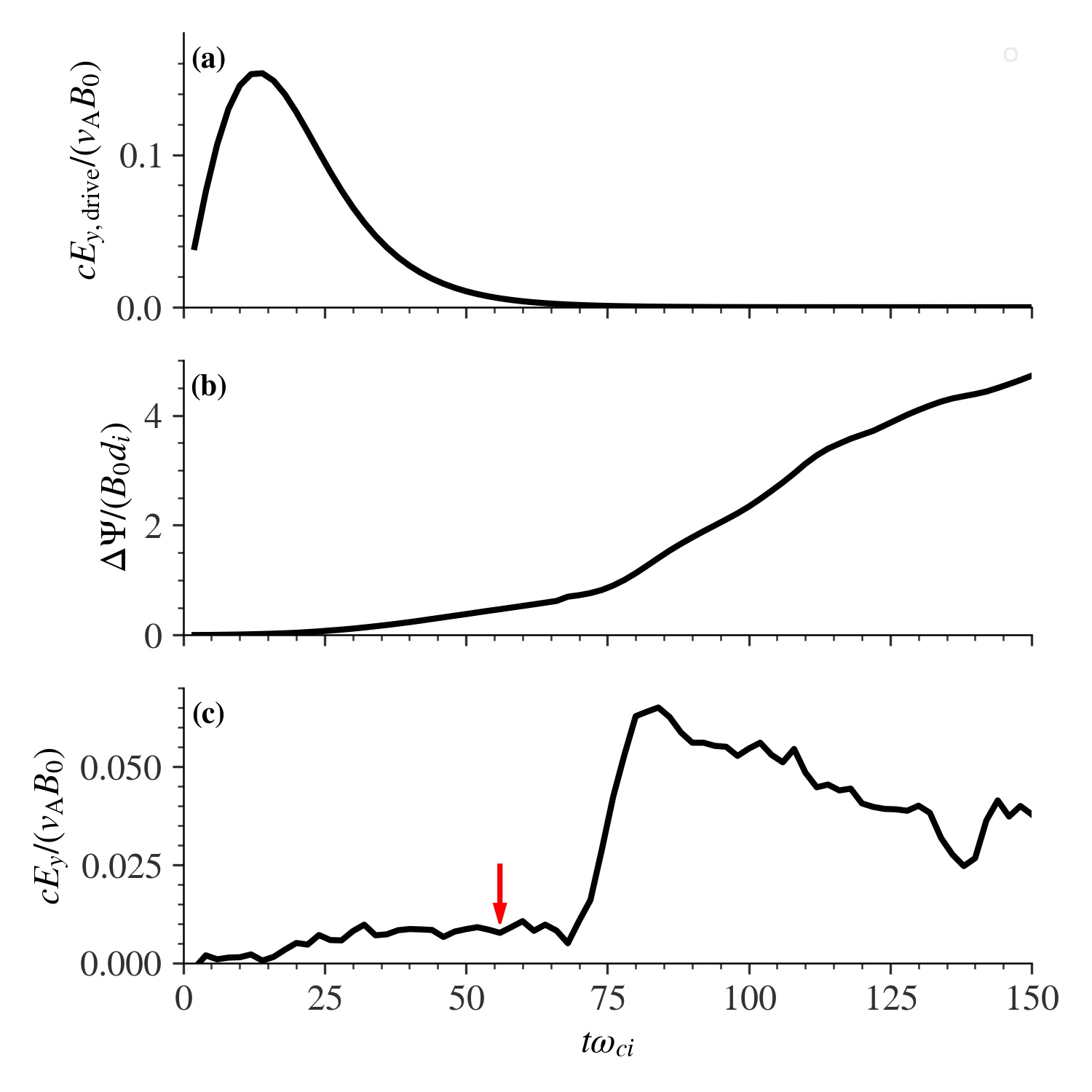}
    \caption{Overview of the simulated magnetic reconnection in a polarized Lembege-Pellat current sheet. (a) The time profile of external electric field $E_y$ applied at the $z$ boundaries. (b) The reconnected magnetic flux $\Delta \Psi$ as a function of time. Before reconnection, $\Delta \Psi$ is calculated as $A_y(x = -L_x) - A_y(x = -L_x/2)$; After reconnection, $\Delta \Psi$ is calculated as $A_y(x = -L_x) - A_y(x = x_\mathrm{rec})$. Here $A_y$ is the $y$ component of magnetic vector potential and $x_\mathrm{rec}$ is the $x$ location of the reconnection site. (c) The reconnection electric field $E_y$ as a function of time. Before reconnection, $E_y$ is taken at $(x = -L_x/2, z = 0)$; After reconnection $E_y$ is taken at $(x = x_\mathrm{rec}, z = 0)$. The red arrow marks the time instant that we study in detail below.}
    \label{fig:reconnection-rate}
\end{figure*}

An overview of the simulation is shown in Figure \ref{fig:reconnection-rate}. The electric field $E_y$ applied at the $z$ boundaries [Figure \ref{fig:reconnection-rate}(a)] drives plasma inflow and adds magnetic flux into the lobe. During this loading process, the current sheet thins and the associated peak current density increases to $\sim e n_0 v_\mathrm{A}$ right before reconnection. At $t = 70\, \omega_{ci}^{-1}$, magnetic field lines reconnect in the plane $z = 0$. Figures \ref{fig:reconnection-rate}(b) and \ref{fig:reconnection-rate}(c) show the time histories of the reconnected magnetic flux and reconnection electric field, respectively. The maximum reconnection rate reaches $0.06$, which is considered as fast reconnection \cite{Birn01,bhattacharjee2009fast}.

\subsection{Force balance in thin current sheet}
\begin{figure*}[tphb]
    \centering
    \includegraphics[width=\textwidth]{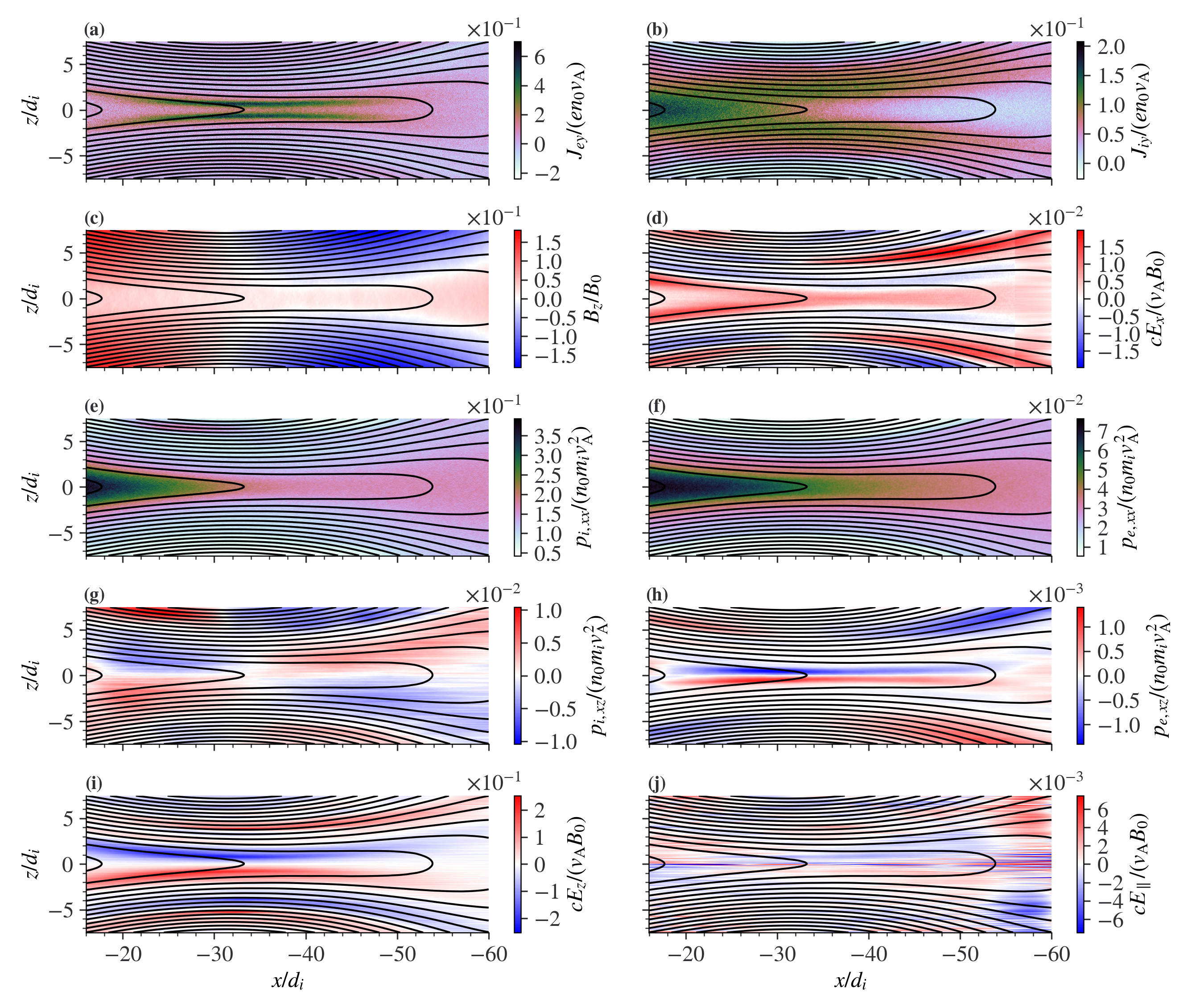}
    \caption{Key physical quantities in the force balance of the thin current sheet. In this example, the two-dimensional maps presented are shown for the time of the red arrow in Figure \ref{fig:reconnection-rate}(c), $t = 56\, \omega_{ci}^{-1}$, shortly before the reconnection onset. (a) The electron current density $J_{ey}$. (b) The ion current density $J_{iy}$. (c) The magnetic field in the $z$ direction $B_z$. (d) The electric field in the $x$ direction. (e) The diagonal ion pressure $P_{i, xx}$. (f) The diagonal electron pressure $P_{e, xx}$. (g) The nongyrotropic (off-diagonal) ion pressure $P_{i, xz}$. (h) The nongyrotropic electron pressure $P_{e, xz}$. \blue{(i) The electric field in the $z$ direction. (j) The electric field parallel to the magnetic field.}}
    \label{fig:current-pressure}
\end{figure*}

\begin{figure*}[tphb]
    \centering
    \includegraphics[width=\textwidth]{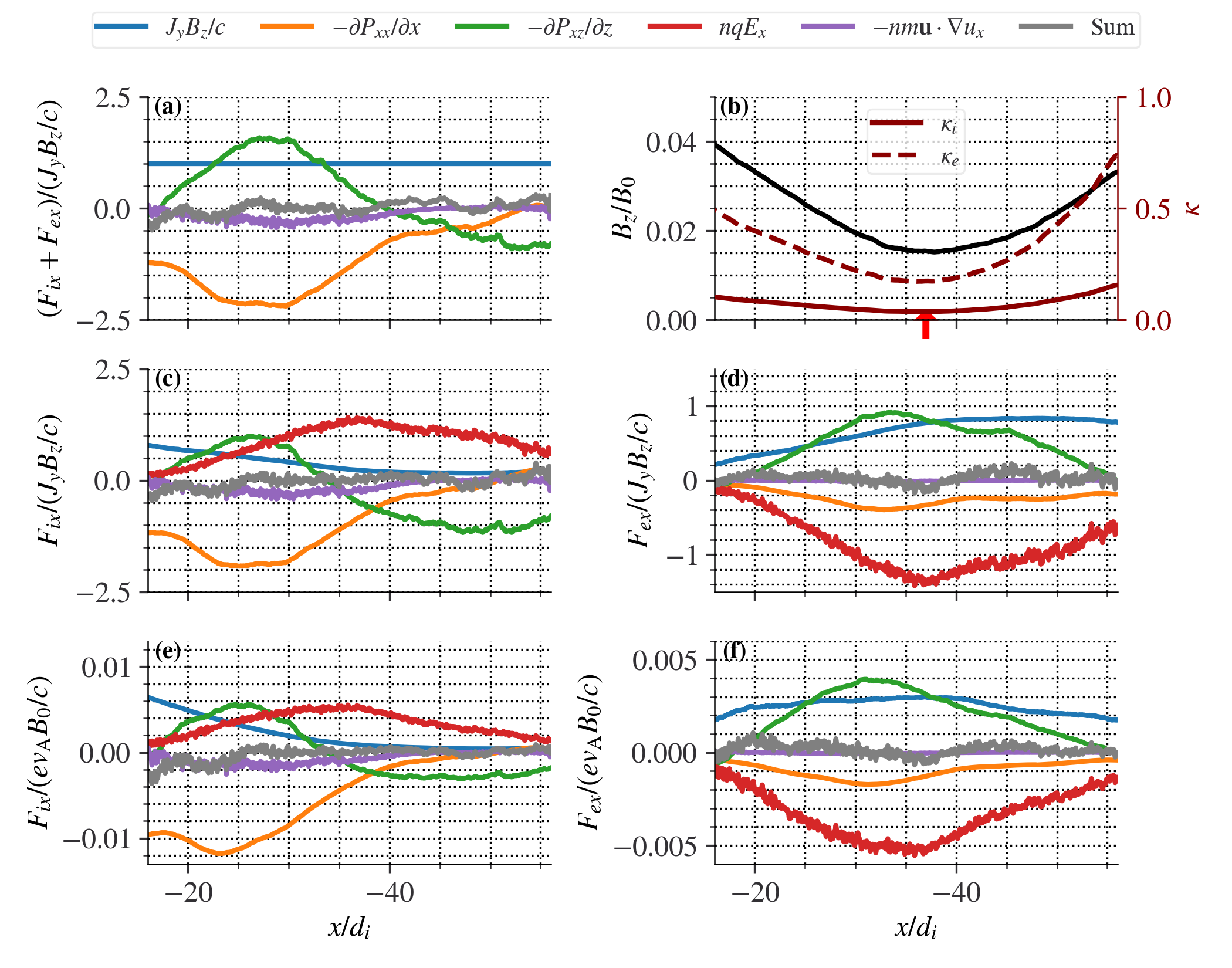}
    \caption{Force balance of the thin current sheet. We consider the balance of volume forces along the $x$ direction in the equatorial plane $z=0$ at $t = 56\, \omega_{ci}^{-1}$. The five terms from the momentum equation, $J_{\alpha y} B_z / c$, $-\partial P_{\alpha, xx} / \partial x$, $-\partial P_{\alpha, xz} / \partial z$, $n q_\alpha E_x$, and $-n m_\alpha \mathbf{u}_\alpha \cdot \nabla u_{\alpha x}$, are represented by blue, orange, green, red, and purple lines, respectively. The sum of these five terms, denoted by gray lines for each species and their sum, is close to $0$, indicating force balance. All the terms in (a), (c) and (d) are normalized by the total magnetic tension force $J_{y} B_z /c$. (a) Force balance from the single-fluid perspective. (b) The profile of $B_z$ \blue{and the adiabaticity parameters $\kappa_{i, e}$} at $z=0$. \blue{The arrow points to the location that will become an active reconnection site later.} (c), (d) Force balance of ions and electrons from the two-fluid perspective. \blue{(e), (f) The same as panels (c) and (d), but in absolute values.}}
    \label{fig:force-balance}
\end{figure*}

We are interested in the configuration of the thin current sheet right before the reconnection onset. The formation of such a thin current sheet is evident from Figures \ref{fig:current-pressure}(a) and \ref{fig:current-pressure}(b). The electron-dominant current sheet bifurcates into two layers \cite<e.g.,>[]{Runov03,Sitnov03}, each with a thickness of $\sim 0.5\, d_i$ (full width at half $J_y$ maximum). We study the force balance of this configuration in the equatorial plane ($z=0$). Figure \ref{fig:force-balance} shows how the nongyrotropic pressure $P_{xz}$ alters the picture of force balance from the initial equilibrium (see the evolution of different terms in the momentum equation at \url{https://doi.org/10.5281/zenodo.6156922}). In the initial equilibrium, from the single-fluid viewpoint, the magnetic tension force at $z=0$ is balanced by the gradient of isotropic plasma pressure, i.e., $J_{y} B_z /c = \partial P_{xx} / \partial x$. In contrast, right before the reconnection onset, there is the nongyrotropic pressure gradient $-\partial P_{xz} / \partial z$ that changes direction around $x / d_i = -40$. This is demonstrated in Figure \ref{fig:force-balance}(a) which shows the forces contributed by isotropic and anisotropic pressure gradients, inertia, and the Lorenz force, as well as their sum, normalized to the Lorenz force and plotted as a function of $x$. In the range $x / d_i > -40$, both $J_{y} B_z /c$ and $-\partial P_{xz} / \partial z$ point in the $+x$ direction, which are balanced by $-\partial P_{xx} / \partial x$ pointing in the $-x$ direction. Around $x / d_i = -40$, the magnetic tension force $J_yB_z/c$ is mainly balanced by the diagonal pressure gradient $-\partial P_{xx} / \partial x$, whereas the off-diagonal pressure gradient $-\partial P_{xz} / \partial z$ is approximately $0$. In the range $x / d_i < -50$, the current sheet becomes almost one dimensional ($\vert \partial P_{xx} / \partial x \vert \ll \vert \partial P_{xz} / \partial z \vert$), where $J_y B_z /c$ is mainly balanced by $-\partial P_{xz} / \partial z$. It is worthwhile to note that the magnetic reconnection later occurs in the transition region around $x / d_i = -40$ [see the profile of $B_z$ in Figures \ref{fig:force-balance}(b) and \ref{fig:current-pressure}(c)].

From the two-fluid approximation, the detailed force balance at $z=0$ in the slowly evolving state ($\partial / \partial t \approx 0$) may be written as
\begin{linenomath*}
\begin{eqnarray}
  -n m_\alpha \mathbf{u}_\alpha \cdot \nabla u_{\alpha x} + \frac{J_{\alpha y} B_z}{c} + n q_\alpha E_x - \frac{\partial P_{\alpha, xz}}{\partial z} - \frac{\partial P_{\alpha, xx}}{\partial x} = 0 ,\,\,\, \text{with }\alpha = e, i .
\end{eqnarray}
\end{linenomath*}
These terms are shown in Figures \ref{fig:force-balance}(c) and \ref{fig:force-balance}(d). The transport terms $-n m_\alpha \mathbf{u}_\alpha \cdot \nabla u_{\alpha x}$ for both ions and electrons are small relative to other terms. The magnetic tension force is dominantly contributed by the electron current, because the current sheet is electron-dominant [as evidenced by looking at Figures \ref{fig:current-pressure}(a) and \ref{fig:current-pressure}(b)]. \blue{The ion adiabaticity parameter $\kappa_i < 0.1$ \cite<$\kappa$ being the square root of the ratio of magnetic field line curvature radius to particle thermal gyroradius at $B_x=0$; see>[]{BZ89} indicates that ions are mostly demagnetized [Figure \ref{fig:force-balance}(b)]. Such ions likely move along quasi-adiabatic orbits and thus contribute significantly to the pressure nongyrotropy \cite<see discussion in>[]{Maha94,Sitnov00,Zelenyi00}. The electron adiabaticity parameter $\kappa_e \sim 0.2-0.3$ indicates stochastic electron motion and strong violation of adiabatic invariant \cite<see review by>[and references therein]{Zelenyi13:UFN}. Difference of ion and electron dynamics leads to current sheet polarization, i.e., ions and electrons are coupled through the electric field $E_x$ [Figure \ref{fig:current-pressure}(d)]. Although $E_x$ is weaker than $E_z$ [Figure \ref{fig:current-pressure}(i)], the entire drift in the neutral plane (where $B_x=0$ and $E_z=0$) is determined by $E_x$, which controls the redistribution of ion and electron contributions to the current density \cite<e.g.,>[]{Zelenyi10GRL}. Note due to the weakness of field-aligned electric fields [Figure \ref{fig:current-pressure}(j)], $E_x\approx -E_zB_z/B_x$.} The diagonal pressure gradient $-\partial P_{xx} / \partial x$ is mostly contributed by ions because of the larger ion temperature, $P_{i,xx} \gg P_{e,xx}$ [Figures \ref{fig:current-pressure}(e) and \ref{fig:current-pressure}(f)].  Although $P_{e, xz}$ is small relative to $P_{i, xz}$ (most likely due to the same temperature ordering), the magnitude of $\vert \partial P_{e, xz} /\partial z \vert$ is comparable to $\vert \partial P_{i, xz} /\partial z \vert$, because the scale length of $P_{e, xz}$ in the $z$ direction is small compared to that of $P_{i, xz}$ [Figures \ref{fig:current-pressure}(g) and \ref{fig:current-pressure}(h)]. The nongyrotropic ion pressure has a quadrupolar structure in the $x$-$z$ plane. The gradient $-\partial P_{i, xz} / \partial z$ both above and below the neutral sheet ($\vert z / d_i \vert < 2$) changes its orientation from pointing in the $+x$ direction on one side $(x / d_i > -35)$ to pointing in the $-x$ direction on the other side $(x / d_i < -35)$ [Figure \ref{fig:current-pressure}(g)]. This is also seen in Figure \ref{fig:force-balance}(c), which shows the force balance in the $x$-direction of the ion fluid, coupled to the electron fluid via the electric force. Figure \ref{fig:force-balance}(d), which shows the same as \ref{fig:force-balance}(c) but for electron fluid, shows that the electron anisotropic pressure gradient is comparable to that of the ions. As we will demonstrate below, the quadrupole of $P_{i, xz}$ is intimately related to the ion flow pattern driven at the boundaries, in which $(x / d_i = -35,\, z = 0)$ is near the inflow stagnation point. 

As can be seen from Figures \ref{fig:force-balance}(c) and \ref{fig:force-balance}(d), on the earthward side $x / d_i > -35$, both $- \partial P_{i, xz} / \partial z$ and $- \partial P_{e, xz} / \partial z$ act in the same direction as $J_{y} B_z / c$, leading to the balance $J_{y} B_z / c - \partial P_{xz} / \partial z  \approx  \partial P_{xx} / \partial x$. Thus, in this region the pressure nongyrotropy effectively decreases the current density ($J_y<c(\partial P_{xx}/\partial x)/B_z$) and makes the current sheet thicker. In the transition region around $x / d_i = -40$ where magnetic reconnection will occur, $- \partial P_{i, xz} / \partial z$ approximately cancels $- \partial P_{e, xz} / \partial z$. The balance in this transition region may be written as $J_{y} B_z / c \approx \partial P_{xx} / \partial x$. On the tailward side $x / d_i < -50$, the nongyrotropic ion pressure gradient becomes dominant, providing the balance $J_{y} B_z / c \approx \partial P_{i, xz} / \partial z$. Thus, in this region the pressure nongyrotropy increases the current density ($J_y>c(\partial P_{xx}/\partial x)/B_z$) and thins the current sheet.

Due to the presence of nongyrotropic pressure gradients, the configuration of thin current sheets just prior to the reconnection onset in the magnetotail is very different from those current sheets with $J_{y} B_z / c \approx \partial P_{xx} / \partial x$ at geomagnetically quiet times. Note even in the region with $\partial P_{i, xz} / \partial z \approx -\partial P_{e, xz} / \partial z$ and $J_y\approx c(\partial P_{xx}/\partial x)/B_z$, where the magnetic reconnection takes place, the current sheet configuration is very different from the initial isotropic 2D equilibrium. In the isotropic 2D equilibrium, any current density increase, which is important to drive the instability leading to the reconnection onset, requires change of large-scale gradient $\partial P_{xx}/\partial x$ or $B_z$ magnitude. In contrast, in the new configuration, any perturbations of the fine balance $\partial P_{i, xz} / \partial z \approx -\partial P_{e, xz} / \partial z$ will drive the current density increase ($\partial P_{i, xz} / \partial z > -\partial P_{e, xz} / \partial z$) or decrease ($\partial P_{i, xz} / \partial z < -\partial P_{e, xz} / \partial z$) for balance to be maintained. Therefore, in such a system even small changes in the pressure nongyrotropy can easily provide a mechanism for further current density increase and possible current sheet destabilization, or further topological reconfiguration and subsequent instability. The stability or instability of such thin current sheet with $\partial P_{i, xz} / \partial z, \partial P_{e, xz} / \partial z \neq  0$ is critical to our understanding of the magnetospheric substorm onset.

\subsection{Origin of pressure nongyrotropy}
To find the origin of the nongyrotropic pressure $P_{\alpha, xz}$ (the subscript $\alpha$ is dropped below for convenience), we multiply the Vlasov equation by the second-order velocity moment $m \mathbf{vv}$, integrate the product in velocity space and simplify the resulting equation using the mass continuity and momentum equations \cite<e.g.,>[]{hakim2008extended}. The final equation governing the evolution of $P_{xz}$ reads
\begin{linenomath*}
\begin{equation}
	\begin{split}
		\frac{\partial P_{xz}}{\partial t} =& \underbrace{-\mathbf{u} \cdot \nabla P_{xz} - 2 \left(\nabla \cdot \mathbf{u}\right) P_{xz} -\frac{\partial u_x}{\partial z} P_{zz} - \frac{\partial u_z}{\partial x} P_{xx}}_{\text{Flow motion } \mathcal{F}_\mathbf{u} (\mathbf{P})} \\
		&\underbrace{+\omega_c \left(b_z P_{yz} + b_y P_{xx} - b_y P_{zz} - b_x P_{xy}\right)}_{\text{Magnetic transformation } \mathcal{M}_\mathbf{B}(\mathbf{P})} \\
        &\underbrace{+ \frac{\partial Q_{xzx}}{\partial x} + \frac{\partial Q_{xzz}}{\partial z}}_{\text{Divergence of heat flux}},
	\end{split}
\end{equation}
\end{linenomath*}
where $\omega_c = B_0 q / (m c)$ is the cyclotron frequency and $\mathbf{b} = \mathbf{B} / B_0$ is the normalized magnetic field, respectively. The local growth of the nongyrotropic pressure $P_{xz}$ can be induced by the flow motion $\mathcal{F}_\mathbf{u} (\mathbf{P})$, the magnetic transformation $\mathcal{M}_\mathbf{B}(\mathbf{P})$, or the divergence of the heat flux tensor. Specifically, $\mathcal{F}_\mathbf{u} (\mathbf{P})$ includes terms due to flow transport ($-\mathbf{u} \cdot \nabla P_{xz}$), flow compression ($- 2 \left(\nabla \cdot \mathbf{u}\right) P_{xz}$) and flow shears ($-\frac{\partial u_x}{\partial z} P_{zz} - \frac{\partial u_z}{\partial x} P_{xx}$); $\mathcal{M}_\mathbf{B}(\mathbf{P})$ contains all possible transformations of other pressure tensor elements into $P_{xz}$ due to the magnetic field.

\begin{figure*}[tphb]
    \centering
    \includegraphics[width=\textwidth]{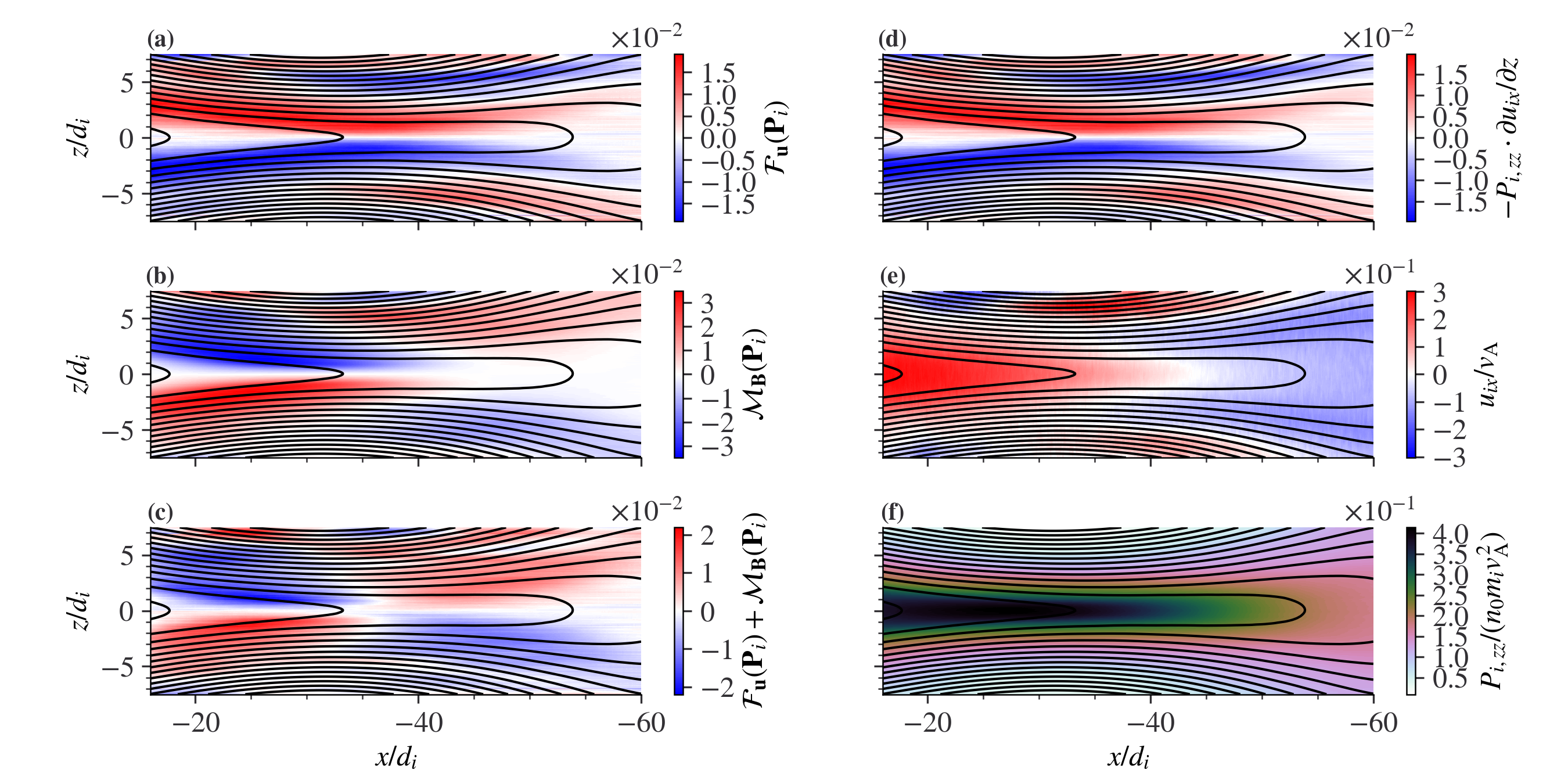}
    \caption{Origin of nongyrotropic ion pressure $P_{i, xz}$. The two-dimensional maps are at $t = 56\, \omega_{ci}^{-1}$. (a) The growth rate of $P_{i, xz}$ due to the flow motion $\mathcal{F}_\mathbf{u} (\mathbf{P})$. (b) The growth rate of $P_{i, xz}$ due to the magnetic transformation $\mathcal{M}_\mathbf{B}(\mathbf{P})$. (c) The total growth rate of $P_{i, xz}$ resulting from the sum $\mathcal{F}_\mathbf{u} (\mathbf{P}) + \mathcal{M}_\mathbf{B}(\mathbf{P})$. (d) The contribution of ion flow shear $-P_{i, zz} \cdot \partial u_{ix} /\partial z$ to the generation of $P_{i, xz}$. The color map of this flow shear term is nearly identical to panel (a). (e) The ion flow in the $x$ direction $u_{ix}$. (f) The diagonal ion pressure $P_{i, zz}$.}
    \label{fig:pxz-generation}
\end{figure*}

For ion pressure nongyrotropy, Figures \ref{fig:pxz-generation}(c) and \ref{fig:current-pressure}(e) show that the local growth rate $\mathcal{F}_\mathbf{u} (\mathbf{P}_i) + \mathcal{M}_\mathbf{B}(\mathbf{P}_i)$ indeed gives rise to the quadrupolar distribution of $P_{i, xz}$. \blue{Note that the divergence of ion heat flux is small compared to $\mathcal{F}_\mathbf{u} (\mathbf{P}_i)$ and $\mathcal{M}_\mathbf{B}(\mathbf{P}_i)$.} Furthermore, comparing Figures \ref{fig:pxz-generation}(a), \ref{fig:pxz-generation}(b) and \ref{fig:pxz-generation}(c), it is clear that the magnetic transformation $\mathcal{M}_\mathbf{B}(\mathbf{P}_i)$ controls the growth of $P_{i, xz}$ on the earthward side $(x / d_i > -35)$ whereas the flow motion $\mathcal{F}_\mathbf{u} (\mathbf{P}_i)$ dominates the $P_{i, xz}$ growth on the tailward side $(x / d_i < -35)$. The growth rate resulting from the flow motion $\mathcal{F}_\mathbf{u} (\mathbf{P}_i)$ is almost all contributed by the ion flow shear $\partial u_{ix} /\partial z$ [Figures \ref{fig:pxz-generation}(a) and \ref{fig:pxz-generation}(d)]. Since the tailward ion flow is maximized off the equator [Figure \ref{fig:pxz-generation}(e)], the velocity shear of this tailward ion flow generates the nongyrotropic pressure $P_{i,xz}$. The gradient $-\partial P_{i, xz} / \partial z$ in turn balances part of the magnetic tension force on the tailward side $(x / d_i < -35)$. In other words, it is the velocity shear by the off-equatorial, tailward ion flow that plays a critical role in the generation of pressure nongyrotropy $P_{i, xz}$ and further maintains the configuration of the thin current sheet through $J_y B_z / c \approx \partial P_{xz} / \partial z + \partial P_{xx} / \partial x>\partial P_{xx} / \partial x$. Interestingly, such velocity shear (although occurring in other boundary layers, e.g., magnetopause, plasma sheet boundary layer) has been previously suggested as the cause of pressure nongyrotropy/anisotropy \cite{de2016pressure,del2016pressure,del2018shear}, which can further generate micro-instabilities in velocity space.

Investigating the generation of electron pressure nongyrotropy, we find that neither the flow motion $\mathcal{F}_\mathbf{u} (\mathbf{P}_e)$ nor the magnetic transformation $\mathcal{M}_\mathbf{B}(\mathbf{P}_e)$ can correctly account for the magnitude and polarity of $P_{e, xz}$ (not shown). Thus, the divergence of the heat flux tensor most likely generates $P_{e, xz}$ in a thin layer [the transverse scale of which being the local electron inertial length; see Figure \ref{fig:current-pressure}(h)] near the equator. Because the heat flux tensor is a high-order velocity moment, this implies that the generation of $P_{e, xz}$ is intrinsically a kinetic process. The critical role of electron heat flux in determining the nongyrotropic electron pressure has been discovered in the electron current layer of magnetic reconnection \cite{Hesse04,Aunai13}. In our simulation the electron demagnitization, resulting in formation of the heat flux divergence and the pressure nongyrotropy, may be partially contributed by an unrealistically large electron mass. Therefore, an additional simulation with low noise (allowing the calculation of the heat flux divergence) is required to address the electron contribution to the pressure nongyrotropy. \blue{The contribution of ion nongyrotropy to the current sheet configuration, however, is much more reliable, because ion demagnitization is quite natural for thin current sheets \cite{Maha94} and the simulation reveals the mechanism for $P_{i,xz}$ formation.}     

\section{Summary and discussion}

In summary, we demonstrate that the configuration of thin current sheets in the magnetotail right before reconnection onset differs significantly from the most investigated 2D equilibrium with $J_y B_z / c \approx \partial P_{xx} / \partial x$. During the current sheet thinning driven by external plasma inflow, the ion flow pattern includes a tailward ($u_{ix}$) and an equatorward ($u_{iz}$) flow \cite<see also>[and references therein]{Sitnov21:grl}. The associated flow shear $\partial u_{ix} /\partial z$ drives the formation of the \blue{ion pressure nongyrotropy, $P_{i,xz}$,} the key element of 1D current sheet models with $B_z\ne 0$ \cite<see>[and references therein]{Zelenyi11PPR,Sitnov&Merkin16}. A finite $P_{xz}$ may explain the formation of current sheets with $J_y$ exceeding the isotropic limit $c(\partial P_{xx} / \partial x)/B_z$ \cite<see discussion of such strong $J_y$ observations in the magnetotail in>[]{Artemyev21:grl}.

In our simulation, the nongyrotropic pressure gradient $-\partial P_{xz} / \partial z$ reverses direction near the inflow stagnation point. This gradient is oriented in the same (opposite) direction as $J_y B_z / c$ on the earthward (tailward) side of the stagnation point. Therefore, there exists a transition of current sheet configurations from a sub-isotropic current ($J_y<c(\partial P_{xx} / \partial x)/B_z$) on the earthward side to a strong current exceeding the isotropic limit ($J_y>c(\partial P_{xx} / \partial x)/B_z$) on the tailward side, with the latter configuration being mostly 1D. It is in this transition region, where $-\partial P_{i, xz} / \partial z$ approximately cancels $-\partial P_{e, xz} / \partial z$, and where magnetic reconnection later occurs. These results underline the importance of accounting for kinetic effects ($P_{xz}\ne0$) in analysis of current sheet instabilities triggering the reconnection onset. 

\section{Open Research}
The data that support the finding of this study have been archived at \url{https://doi.org/10.5281/zenodo.6156922}.

\acknowledgments
The work was supported by NASA awards 80NSSC18K1122, 80NSSC20K1788, and NAS5-02099. We would like to acknowledge high-performance computing support from Cheyenne (\url{doi:10.5065/D6RX99HX}) provided by NCAR's Computational and Information Systems Laboratory, sponsored by the National Science Foundation.


%
%


%
%
%
%
%

\end{document}